\documentstyle[aps,prl,preprint,epsfig]{revtex}

\def\begt{\begin{tabular}}
\def\endt{\end{tabular}}
\def\bege{\begin{equation}}
\def\ende{\end{equation}}

\begin{document}
\title{Quantum shock waves in the Heisenberg XY model}

\author{Vladislav Popkov
\footnote{also at the Institute for Low Temperature Physics,
Kharkov, Ukraine.\\
E-mail: popkov@physik.fu-berlin.de}\\
\small{Institut f\"ur Theoretische Physik, Freie Universit\"at Berlin,\\
Arnimallee 14, 14195 Berlin, Germany}\\
and\\
Mario Salerno
\footnote{also at the Istituto Nazionale di Fisica
della Materia (INFM) Unit\`a di Salerno.\\
E-mail: salerno@sa.infn.it}\\ 
\small{Dipartimento di Scienze Fisiche "E.R. Caianiello",\\
via S. Allende, I-84081 Baronissi (SA), Italy.}}
\maketitle

\begin{abstract}
We show the existence of quantum states of the Heisenberg XY chain which
closely follow the motion of the corresponding semi-classical ones, and
whose evolution resemble the propagation of a shock wave in a fluid. These
states are exact solutions of the Schr\"odinger equation of the XY model and
their classical counterpart are simply domain walls or soliton-like
solutions.

\end{abstract}

\newpage

\section{Introduction}

The relationship between classical and quantum dynamics, although largely
investigated since the beginning of quantum mechanics, is still a non
exhausted problem of ever continuing interest. This has lead, for example,
to the introduction of new concepts such as coherent states and their
generalizations, which has been found to be successful in describing the
quasiclassical behaviors of a wide class of physical problems. The
possibility to construct quantum states which closely resemble the evolution
of classical ones is appealing also in connection with the new developing
area of quantum computing \cite{quantum_computer}. In this context the basic
component of a quantum computer is a q-bit i.e. a special quantum state
which has to be manipulated by the logical quantum gates (unitary operators)
and which has to be preserved `intact' for an extended period of time. In
any practical task, however, the interaction of the system with the
environment (heat reservoir) is unavoidable and decoherence usually
develops. On the other hand, one may expect that, due to their intrinsic
collective character, quantum states close to macroscopic (classical) ones
are more robust against decoherence and therefore more suitable to store
information in a quantum computer. The problem of finding such states,
however, is far from being trivial even for isolated systems. Indeed, the
complexity of a quantum system related to the huge dimensionality of the
Hilbert space, represents a major obstacle in solving the time-dependent
Schr\"{o}dinger equation (one usually resorts to approximations which are
not always well controlled).

The aim of the present paper is to show that for the XY Heisenberg spin
chain \cite{spinmodel} it is possible to find quantum states which closely
follow the motion of the corresponding classical ones and whose evolution
resemble the propagation of a shock wave in a fluid. The classical
counterpart of these states are moving domain walls or soliton-like
solutions. Stable shock-wave solutions of different shapes were also
encountered in the semiclassical dynamics of the Heisenberg chain \cite
{Mario_hei} as well as in other similar models \cite{Mario_two_level}. Here
we show, on the example of the solvable XY model, that a spin system may
indeed have exact quantum states which recall classical moving shock-wave
solutions. In particular, a detailed comparison between the quantum and the
classical evolution for several initial conditions is presented. The regions
where the exact quantum states are well approximated by the classical ones
are also investigated. In these regions the classical approach can be used
to track the dynamics of those quantum observables which are technically
unaccessible by a quantum-mechanical analysis. The paper is organized as
follows. First we derive semiclassical and quantum equations of motion, and
in particular solve them in the quantum case. This enables us to compare
exact quantum and semiclassical time evolutions for a chosen set of initial
conditions leading to shock-wave formation. The properties of the latter are
then discussed, and in the conclusion the main results of the paper are
summarized.

\section{Semiclassical analysis of the XY model}

The quantum Heisenberg $XY$ model \footnote{%
An arbitrary constant magnetic field $B \sum_{n}s_n^z $ can be included;
here we consider $B=0$ for simplicity.} is introduced as 
\begin{equation}
{\cal H}=J\sum_{n}(s_{n}^{x}s_{n+1}^{x}+s_{n}^{y}s_{n+1}^{y})  \label{XX0}
\end{equation}
where $s^{i}={\frac{1}{2}}\sigma ^{i},\ i=1,2,3$ are spin $1/2$ operators, $%
\sigma ^{i}$ are Pauli matrices, and $J\;$ is the exchange constant related
to the interaction between spins. As well known, the two-dimensional
character of the interaction and its restriction to neighbouring spins leads
to the exact solvability of this model \cite{lieb}. In the following we
shall derive semiclassical equations of motion of the $XY$ model, by
averaging the Heisenberg equation of motion $i\hbar {\frac{\partial \hat{f}}{%
\partial t}}=[\hat{f},{\cal H}]$ for a generic observable $\hat{f}$ over
(uncorrelated) coherent states. For $s_{n}^{z}$ (site magnetization) and $%
s^{\pm }=s^{x}\pm is^{y}$ (rising and lowering operators), we have:
\begin{equation}
i\hbar {\frac{\partial s_{n}^{z}}{\partial t}}={\frac{J}{2}}\left(
s_{n}^{+}s_{n+1}^{-}-s_{n}^{-}s_{n+1}^{+}-s_{n-1}^{+}s_{n}^{-}+s_{n-1}^{-}s_{n}^{+}\right),
\label{s_z}
\end{equation}
\begin{equation}
-i\hbar {\frac{\partial s_{n}^{\pm }}{\partial t}}=\pm Js_{n}^{z}\left(
s_{n+1}^{\pm }+s_{n-1}^{\pm }\right) ,  \label{s_plus}
\end{equation}
(in the following to simplify notation we fix $\hbar =1$ and $J=1$).
Introducing spin ${\frac{1}{2}}$ coherent states as 
\begin{equation}
|\Lambda (t)>=\prod_{n}\otimes |\mu _{n}(t)>,\ \ \ |\mu _{n}(t)>={\frac{%
e^{\mu _{n}(t)s_{n}^{-}}}{\sqrt{1+|\mu _{n}(t)|^{2}}}}|\uparrow \rangle ,
\label{Lambda}
\end{equation}
with $|\uparrow \rangle _{n}$ denoting the spin up state at site $n$, we
compute 
\begin{equation}
\langle s_{n}^{+}\rangle ={\frac{\mu _{n}}{1+|\mu _{n}|^{2}}},\ \ \langle
s_{n}^{-}\rangle ={\frac{\mu _{n}^{*}}{1+|\mu _{n}|^{2}}},\ \ \langle
s_{n}^{z}\rangle ={\frac{1}{2}}\left( \frac{1-|\mu _{n}|^{2}}{1+|\mu
_{n}|^{2}}\right) ,  \label{observables}
\end{equation}
where $\langle s\rangle $ denotes the average over the state $|\Lambda
\rangle $. It is worth noting  that these equations define an inverse
stereographic mapping from the variables $\mu _{n},\mu _{n}^{*}$,
to \ the vectors $\vec{S}_{n}$  (classical spins) on the unit sphere

\begin{equation}
S_{n}^{x}=\frac{\langle s_{n}^{+}\rangle +\langle s_{n}^{-}\rangle }{2}%
,\;S_{n}^{y}=\frac{\langle s_{n}^{+}\rangle +\langle s_{n}^{-}\rangle }{2}%
,\;S_{n}^{z}=\langle s_{n}^{z}\rangle   \label{claspin}
\end{equation}
 By averaging Eqs. (\ref{s_z}), (\ref{s_plus}) with respect to $|\Lambda
\rangle $ and by eliminating redundancy (note that the quantities in Eq. (%
\ref{observables}) are not all independent), we finally obtain the equation
of motion for $\mu _{n}(t)$ as 
\begin{equation}
i{\frac{\partial \mu _{n}}{\partial t}}={\frac{1}{2}}\left( {\frac{\mu
_{n+1}-\mu _{n}^{2}\mu _{n+1}^{*}}{1+|\mu _{n+1}|^{2}}}+{\frac{\mu
_{n-1}-\mu _{n}^{2}\mu _{n-1}^{*}}{1+|\mu _{n-1}|^{2}}}\right).
\label{classical}
\end{equation}
Note that this equation (and its complex conjugated) can be put in
hamiltonian form 
\begin{equation}
\dot{\mu}_{n}=\left\{ \mu _{n},H\right\}   \label{hameq}
\end{equation}
with respect to the Poisson bracket 
\begin{equation}
\{f,g\}=-i\sum_{n}(1+|\mu _{n}|^{2})^{2}\left( \frac{\partial f}{\partial
\mu _{n}}\frac{\partial g}{\partial \mu _{n}^{*}}-\frac{\partial f}{\partial
\mu _{n}^{*}}\frac{\partial g}{\partial \mu _{n}}\right) ,  \label{poisson}
\end{equation}
and the hamiltonian 
\begin{equation}
H_{c}(\mu ,\mu ^{*})=\frac{1}{2}\sum_{n}\frac{\mu _{n+1}^{*}\mu _{n}+\mu
_{n}^{*}\mu _{n+1}}{(1+|\mu _{n}|^{2})(1+|\mu _{n+1}|^{2})}\;.
\label{claham}
\end{equation}
These results were also derived in Ref. \cite{Mario_hei} using the
stationary phase approximation and the path integral formulation of quantum
mechanics. It is remarkable that the classical hamiltonian in
Eq. (\ref{claham}), once expressed in terms of the vectors $\vec{S}_{n}$
through Eqs. (\ref{observables},\ref{claspin}), takes exactly the same
form as in Eq. (\ref{XX0}) with quantum operators replaced by classical
spins. This is precisely what one would expect from the classical
limit in coherent state representation. On the other hand,
since coherent states do not preserve the hidden symmetry algebra
of the quantum XY chain (they are not eigenstates of the monodromy
operator) \cite{lieb-mattis}, the integrability structure of the system is
lost in the semiclassical limit.

\section{Exact quantum analysis of the XY model}

In this section we shall derive exact analytical expressions for the time
evolution
of the expectation values $\langle \hat{f}(t)\rangle =\langle
\psi _{0}|e^{i{\cal H}t}\hat{f}(0)e^{-i{\cal H}t}|\psi _{0}\rangle $ of
observables $\hat{f}(t)$ of the $XY$ model. To this end we take as initial
state $|\psi _{0}\rangle $ the totally uncorrelated quantum state 
\begin{equation}
|\psi _{0}\rangle =\prod_{n=-\infty }^{\infty }\otimes \left( 
\begin{array}{c}
e^{i\Phi _{n}/2}\cos (\alpha _{n}) \\ 
e^{-i\Phi _{n}/2}\sin (\alpha _{n})
\end{array}
\right) .  \label{qstate}
\end{equation}
By the Jordan-Wigner \cite{jord-wig} transformation from spin to
Fermi operators 
\begin{equation}
s_{n}^{+}=\prod_{m<n}\left( 1-2C_{m}^{+}C_{m}\right) C_{n},\ \ \ \ \
s_{n}^{-}=(s_{n}^{+})^{*},\ \ \ \ \ s_{n}^{z}={\frac{1}{2}}-C_{n}^{+}C_{n}, 
\label{J-W}
\end{equation}
the $XY$ Hamiltonian (\ref{XX0}) reduces to
\begin{equation}
{\cal H}={\displaystyle {\frac{1}{2}}}\sum_{n}\left(
C_{n+1}^{+}C_{n}+C_{n}^{+}C_{n+1}\right) .  \label{Hhamilt}
\end{equation}
This Hamiltonian is readily diagonalized as 
\begin{equation}
{\cal H}=\sum_{k}cos(k)c_{k}^{+}c_{k},  \label{diagonal}
\end{equation}
by the Fourier transform $C_{n}={\frac{1}{\sqrt{N}}}\sum_{k}e^{-ikn}c_{k}$.
The Heisenberg equations of motion 
\begin{equation}
i{\frac{\partial c_{k}}{\partial t}}=cos(k)c_{k},
\end{equation}
are then solved as $c_{k}(t)=e^{-itcos(k)}c_{k}(0).$ Returning to the
operators $C_{n}$, we get 
\begin{equation}
C_{n}(t)={\frac{1}{\sqrt{N}}}\sum_{k}e^{-ikn-icos(k)t}c_{k}(0)=%
\sum_{m}i^{m-n}J_{m-n}(-t)C_{m}(0),
\end{equation}
with $J_{n}(t)$ denoting the Bessel function of order $n$. From Eq. (\ref
{J-W}) and from its inverse 
\begin{equation}
C_{n}=\left( \prod_{m<n}s_{m}^{z}\right) s_{n}^{+},\ \ \ \
C_{n}^{+}=s_{n}^{-}\left( \prod_{m<n}s_{m}^{z}\right) ,  \nonumber
\end{equation}
one can finally derive exact expressions for the time evolution of the
observables of the original spin chain. Thus, for example, the time
evolution of the averaged spin-$z$ projection is obtained as 
\begin{eqnarray}
&&S_{p}^{z}(t) \equiv 2\langle s_{p}^{z}\rangle =\langle \psi
_{0}|1-2C_{p}^{+}(t)C_{p}(t)|\psi _{0}\rangle =  \nonumber \\
&&1-2\sum_{n=-\infty }^{\infty }\sum_{m=-\infty }^{\infty
}i^{n-m}J_{n-p}(-t)J_{m-p}(-t)\ \ \langle \psi _{0}|s_{m}^{-}\left(
\prod_{l=-\infty }^{m-1}s_{l}^{z}\right) \left( \prod_{l^{\prime }=-\infty
}^{n-1}s_{l^{\prime }}^{z}\right) s_{n}^{+}|\psi _{0}\rangle =\nonumber \\
&&S_{p}^{z}(0)- \sum_{n<m} J_n(t) J_m(t) sin(2\alpha_{p+n})
sin(2\alpha_{q+n}) 
cos\left( {\pi \over 2}(m-n)+\Phi_{m+p}-\Phi_{n+p} \right)
\prod_{j=n+p+1}^{m+p-1} S_j^z(0)
 \label{quantum}
\end{eqnarray}
(in a similar manner one proceeds for the other observables). In the next
section we shall compare this expression for the average site magnetization,
with direct numerical integrations of Eq. (\ref{classical}).

\section{Comparison between classical and quantum evolutions}

To compare the classical and quantum time evolution
 from an identical initial state,  we shall first derive
the classical initial conditions which correspond to the quantum
state Eq. (\ref{qstate}). To this end we remark that 
the expectation values of the
operators $s_{n}^{\pm },s_{n}^{z}$, with respect to $|\psi _{0}\rangle $ are
easily calculated as 
\begin{equation}
\langle s_{n}^{\pm }\rangle ={\frac {e^{\mp i\Phi_{n}}}{2}}
sin\ 2\alpha _{n},\ \ \ \ \langle s_{n}^{z}\rangle ={\frac{1}{2}}cos\ 2\alpha _{n}.
\end{equation}
The initial conditions in the classical system are fixed by requiring that
the above expressions for $\langle s_{n}^{\pm }\rangle ,\langle s_{n}\rangle
,$ coincide with the corresponding classical ones in Eq. (\ref{observables}%
), this leading to 
\begin{equation}
\mu _{n}(0)=tg(\alpha _{n})e^{-i\Phi _{n}}  \label{icclass}
\end{equation}
as initial conditions for Eq.{\ref{classical}}.
Since the classical states stay uncorrelated for all times while quantum
ones develop correlations in the course of time, one expects the
corresponding classical and quantum time 
evolution to be different for generic
initial conditions. For special choices of the initial conditions however,
these dynamics may qualitatively agree for times long enough to observe
phenomena of the classical system, such as shock waves formation, also in
the quantum one. Shock waves were observed in a semiclassical description of
the Heisenberg chain using as initial conditions non constant (bell shaped)
profiles for $\langle s_{n}^{z}\rangle $ with a constant phase $\Phi
_{n}=const$ along the chain \cite{Mario_hei}. Here it is convenient to take
as initial conditions constant values for the site magnetizations but not
for the phase i.e. we use 
\begin{equation}
S_{n}^{z}(0)=cos(2\alpha ),\;\ \ \Phi _{n}=\Phi (n),\;\;-\frac{N}{2}\leq
n\leq \frac{N}{2}
\end{equation}
with $\alpha $ an arbitrary constant and $N$ the length of the chain. Note
that, except for the case of linearly increasing phases, $\Phi _{n}=\phi n$ (%
$\phi =$constant$)$ for which $S_{n}^{z}$ will stay constant at all times
(this follows directly from Eqs. (\ref{classical},\ref{quantum})), the
inhomogenuity of the phase at $t=0$ will in general induce a spatial
dependence of $S_{z}$ in the course of time. 
In the following we choose the phase of the
classical and quantum initial state so as
 to generate two types of excitations:
localized states and expanding shock wave solutions.

{\bf A) Localized states}.

We have generated these excitations with the following initial conditions

\noindent a) $\Phi _{n}=\phi n,\ n<0;\ \ \Phi _{n}=\phi n+A,\ n\geq 0$; (
dislocation),

\noindent b) $\Phi _{n}=\phi n,\  n\neq 0;\ \ \Phi _{0}=A\neq 0;$ (local
inhomogenuity).

\noindent In Fig. \ref{local_quantum} we depict the time evolution of the
quantum averaged site magnetization as computed from Eq. (\ref{quantum}) for
the state in Eq. (\ref{qstate}) with phase given by a). In Fig.\ref
{local_classic} a direct numerical integration of Eq. (\ref{classical}) with
a fourth order Runge-Kutta method for the corresponding initial condition in
Eq. (\ref{icclass}) and $\Phi _{n}$ given by a), is reported. We see that in
both cases localized soliton-like excitations are formed together with a
radiative field which is much stronger in the case of the classical
evolution. The initial conditions of type b) also lead to the formation of
soliton-like solutions and background radiation, but in this case the
quantum solitons have much shorter life-times and therefore, will not be
discussed here.

{\bf B) Expanding shock-wave solution.}

To form a shock-wave in the magnetization profile we have considered initial
conditions of the form $\Phi _{n}=\phi |n|$, on an infinite open chain (note
that the phase has a cusp at $n=0$). Such a configuration can in principle
be prepared by applying a periodic magnetic field $cos(\omega t)$ to the
left half of the chain and a corresponding $\pi $-shifted field (i.e. $%
-cos(\omega t)$) to the right half. In Figs. \ref{shock_quantum}, \ref
{shock_classic} we depict the time evolution of this initial condition for,
respectively, the quantum and the classical cases. The formation of an
expanding symmetrical shock front is clearly seen in both cases and the
corresponding time propagations are in a good agreement for long times. We
also remark that the expanding inhomogeneous region with higher $S_{n}^{z}$
is caused by the fact that there are two opposite fluxes of magnetization $%
j>0$ (on the left) and $j<0$ (on the right), colliding at the center. At the
initial time $t=0$, they are separated by the single site $n=0$. As time
grows, the opposite magnetization currents become separated by the zero-flux
region $j\approx 0$, expanding alongside with perturbed $S_{n}^{z}$ region
as one can see from Fig.\ref{flux_quantum}. The symmetry of the $S_{n}^{z}$
is just due to the symmetry of the incoming currents.

The relation between the rate of change of magnetization in time and its
current is given by a discrete continuity equation 
\begin{equation}
{\frac{\partial S_{n}^{z}}{\partial t}}=\langle j_{n}\rangle -\langle
j_{n+1}\rangle ,\ \ \ j_{n}=i\left(
s_{n-1}^{+}s_{n}^{-}-s_{n-1}^{-}s_{n}^{+}\right) ,  \label{j_n}
\end{equation}
which is obtained from Eq. (\ref{s_z}) by averaging with respect to the
initial state. Denoting by $-A(t),A(t)$ the boundaries of the perturbed
region, and summing Eq. (\ref{j_n}) over the points inside this region, we
obtain 
\begin{equation}
{\frac{\partial \sum_{-A(t)}^{A(t)}S_{n}^{z}}{\partial t}}%
=\sum_{-A(t)}^{A(t)}\left( \langle j_{n}\rangle -\langle j_{n+1}\rangle
\right) =\langle j_{-A(t)}\rangle -\langle j_{A(t)}\rangle .
\label{sum_S_deriv}
\end{equation}
For the case $\Phi _{n}=\phi |n|$ we find 
\begin{equation}
{\frac{\partial }{\partial t}}\sum_{-A(t)}^{A(t)}S_{n}^{z}=\left(
1-(S^{z}(0))^{2}\right) sin(\phi ),
\end{equation}
from which we see that the excess magnetization in the perturbed $S_{n}^{z}$
region grows linearly in time 
\begin{equation}
\sum_{-A(t)}^{A(t)}\left( S_{n}^{z}(t)-S_{n}^{z}(0)\right) =\alpha t,\ \ \ \
\ \alpha =(1-(S^{z}(0)^{2}))sin(\phi ).  \label{linear_S}
\end{equation}
Using a similar argument for the local energy 
\begin{equation}
E_{n}=\langle s_{n}^{x}s_{n+1}^{x}+s_{n}^{y}s_{n+1}^{y}\rangle ,
\end{equation}
we obtain that the excess energy inside the perturbed region also increases
linearly in time 
\begin{equation}
\sum_{-A(t)}^{A(t)}\left( E_{n}(t)-E_{n}^{z}(0)\right) =\beta t,\ \ \ \beta
=-{\frac{1}{4}}\left( 1-(S^{z}(0))^{2}\right) S^{z}(0)sin(2\phi ).
\label{linear_E}
\end{equation}
This is also seen from Fig.\ref{energy_quantum} where the energy profiles of
a shock solution at different instants of time are reported. It is worth to
note that the magnetization production is maximal (see Eq. (\ref{linear_S})
with $\phi =\pi /2$), when the energy production is zero (see Eq. (\ref
{linear_E})). So, quite surprisingly, the maximal magnetization current is
always accompanied by a zero energy current. We observed this feature also in
quantum evolution of the XXZ spin hamiltonian, for which the time-dependent
Schr\"odinger problem can be exactly solved for special initial conditions 
\cite{ps}. Another interesting feature of the above
solution is its  scaling property
\begin{equation}
S_{n}^{z}(t)\rightarrow S^z(n/t),\ \ 
\ \ n,t \gg 1
  \label{scaling}
\end{equation}
for the site magnetization and energy, as well as for the flux and
the energy $j_n(t), E_{n}(t) $,
which can be proved along the similar lines as in \cite{Gunter}.
Precise form of the scaling function is complicated and is omitted here. 
Finally, we note that Eq. (\ref{scaling})
agrees with Eqs. ({\ref{linear_S}), (\ref{linear_E}), and that the
above scaling properties are  shared ( numerically, at least)
by the classical solution.

\section{Summary}

In this paper we have compared the exact time evolution of the quantum
Heisenberg XY spin 1/2 model with its classical counterpart. In particular,
we have reported on localized states and on expanding shock-wave solutions
whose classical and quantum evolutions agree for long times. The problem of
finding suitable conditions which allow to discriminate the quantum states
which will evolve `quasiclassically' from the ones which will not, remains
however unsolved. Thus, for example, we found that initial conditions
leading to decrease of absolute value of local magnetization (e.g. initial
conditions of Fig.~\ref{shock_classic} with the sign of $\phi $ inverted),
develop fast instabilities in the classical case which are not observed in
the quantum evolution. This could possibly be explained by a difference of
the `dynamical temperature' for the classical spin system in the two cases 
\cite{Schotte_privat}. It will be interesting to extend this analysis to
other initial conditions such as the one used in Ref. \cite{malomed}, as
well as to other quantum systems such as the XXZ model.

\section{Acknowledgments}

The authors thank G.M.Sch\"utz
for useful discussions . Financial support from INFM (Istituto Nazionale di Fisica
della Materia) Sezione di Salerno, Humboldt Foundation and EC grant
LOCNET (contract number HPRN-CT-1999-00163), is acknowledged.

%\begin{thebibliography}{99}

%\end{thebibliography}
\newpage
\section{Figure captions}

%\label{local_quantum}
Fig.1. Snapshots of site spin-z component $s_n^z$ expectation values for
localized excitation  at a different moments of time, plotted from Eq. (\ref
{quantum}).  The initial conditions are: $S_z(0)=0.8, \Phi_n=\phi n, n<0,
\Phi_n=\phi n+2 \phi, n \geq 0; \ \ \phi=\pi/2 $. Graphs $a,b,c,d$
correspond to time=$10,30,50,100$ respectively.

%\label{local_classic}
Fig.2. the same as in Fig.1 but for the semiclassical evolution obtained
from Eq. (\ref{classical}). Graphs $b,c,d$ correspond to time=$30,50,100$
respectively.

%\label{shock_quantum}
Fig.3. Average magnetization profile $\langle s_n^z \rangle $ for shock
solution (exact) at different times, computed from Eq. (\ref{quantum}).
Consecutive expanding curves correspond to $t =10,30,50,100$. The initial
conditions are: $S_z(0)=0.8, \Phi_n=\phi |n|, \ \ \phi=\pi/4 $.

%\label{shock_classic}
Fig.4. the same as in Fig.3 but for the semiclassical evolution.

%\label{flux_quantum}
Fig.5. Flux profile (exact) at time $t =0,30,50,100$. Initial conditions are
the same as in Fig.3.

%\label{energy_quantum}
Fig.6. Energy profile for a shock solution ( $t =0,30,50,100$). Initial
conditions are the same as in Fig.3.

\begin{figure}[htbp]
\setlength{\unitlength}{1.5cm}\vspace{3mm}\hspace*{-2cm}
\epsfig{width=7\unitlength,
%      angle=-90, 
      angle =0,
%      bbllx=250pt, bblly=50pt, bburx=554pt, bbury=770pt,
      file=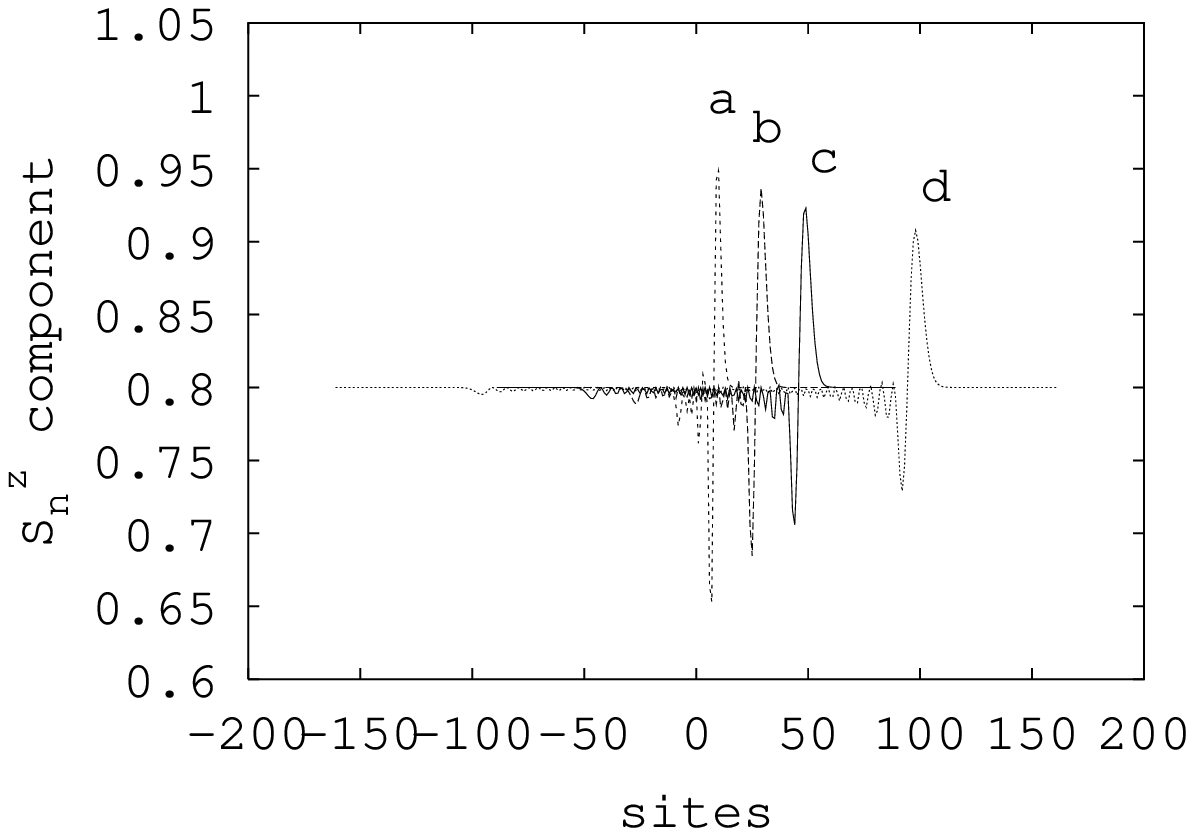}\vspace{3mm}
\caption[]{}
\label{local_quantum}
\end{figure}

\begin{figure}[htbp]
\setlength{\unitlength}{1.5cm}\vspace{3mm}\hspace*{-2cm}
\epsfig{width=7\unitlength,
%      angle=-90, 
      angle =0,
%      bbllx=250pt, bblly=50pt, bburx=554pt, bbury=770pt,
      file=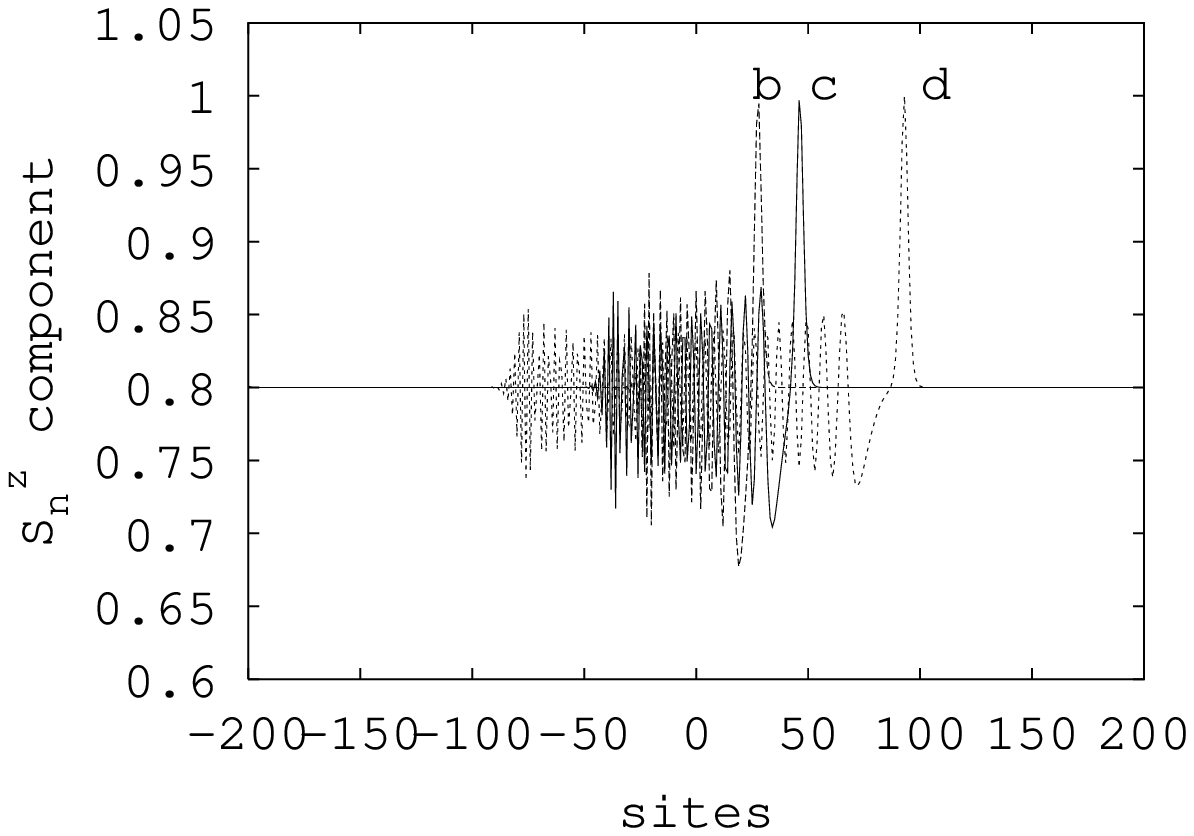}\vspace{3mm}
\caption[]{}
\label{local_classic}
\end{figure}

\begin{figure}[htbp]
\setlength{\unitlength}{1.5cm}\vspace{3mm}\hspace*{-2cm}
\epsfig{width=7\unitlength,
%      angle=-90, 
      angle =0,
%      bbllx=250pt, bblly=50pt, bburx=554pt, bbury=770pt,
      file=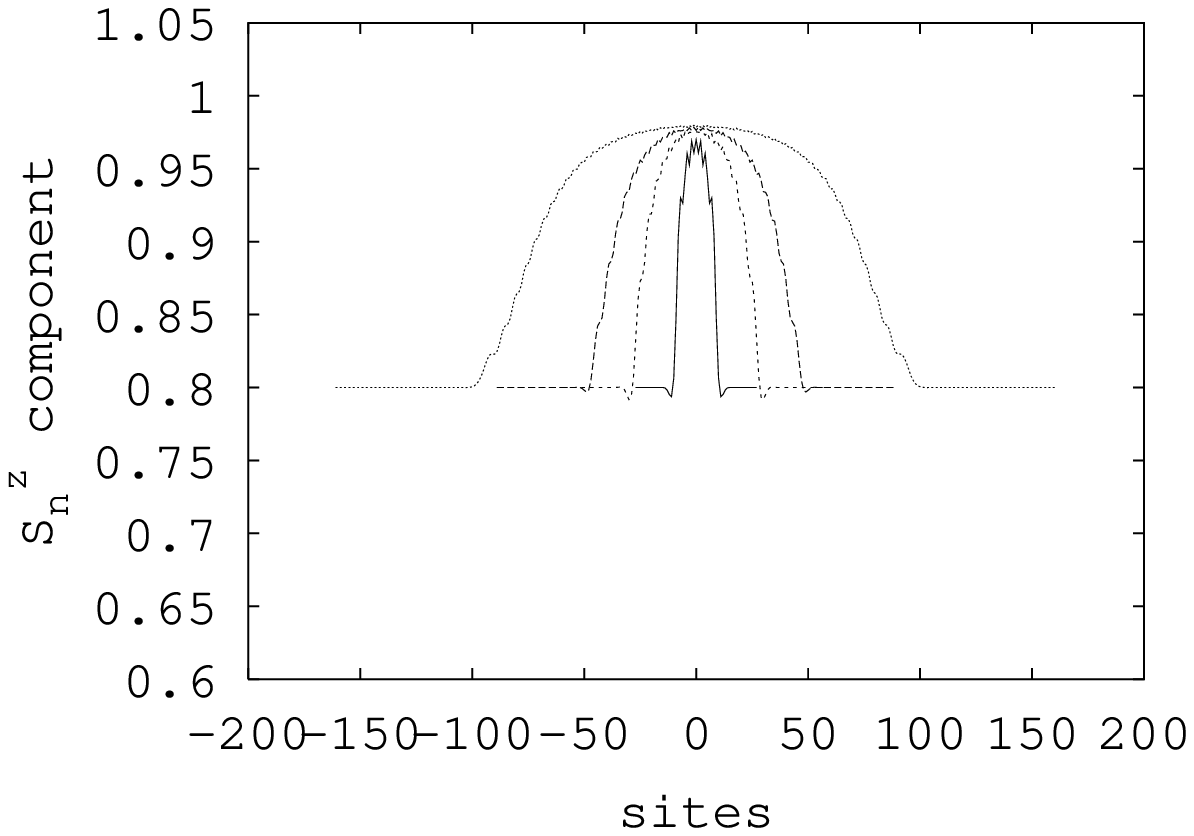}\vspace{3mm}
\caption[]{}
\label{shock_quantum}
\end{figure}

\begin{figure}[htbp]
\setlength{\unitlength}{1.5cm}\vspace{3mm}\hspace*{-2cm}
\epsfig{width=7\unitlength,
%      angle=-90, 
      angle =0,
%      bbllx=250pt, bblly=50pt, bburx=554pt, bbury=770pt,
      file=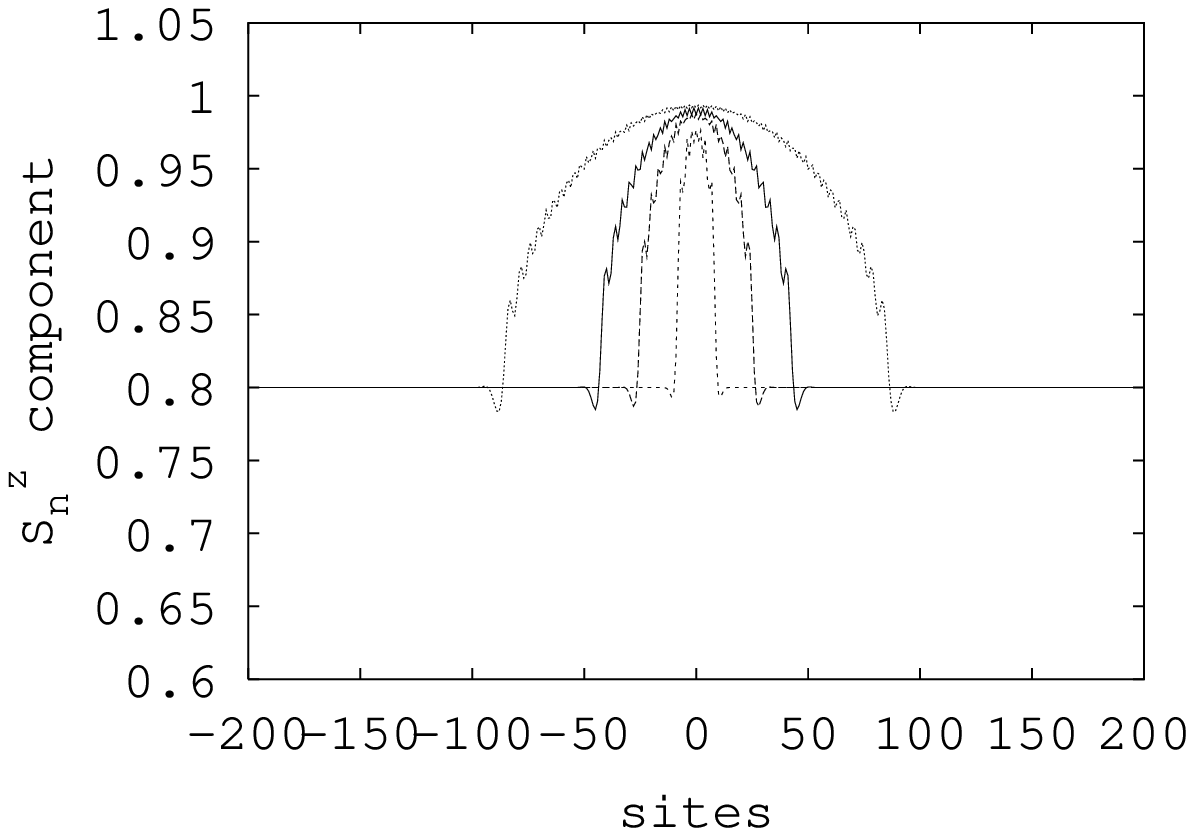}\vspace{3mm}
\caption[]{}
\label{shock_classic}
\end{figure}

\begin{figure}[htbp]
\setlength{\unitlength}{1.5cm}\vspace{3mm}\hspace*{-2cm}
\epsfig{width=7\unitlength,
%      angle=-90, 
      angle =0,
%      bbllx=250pt, bblly=50pt, bburx=554pt, bbury=770pt,
      file=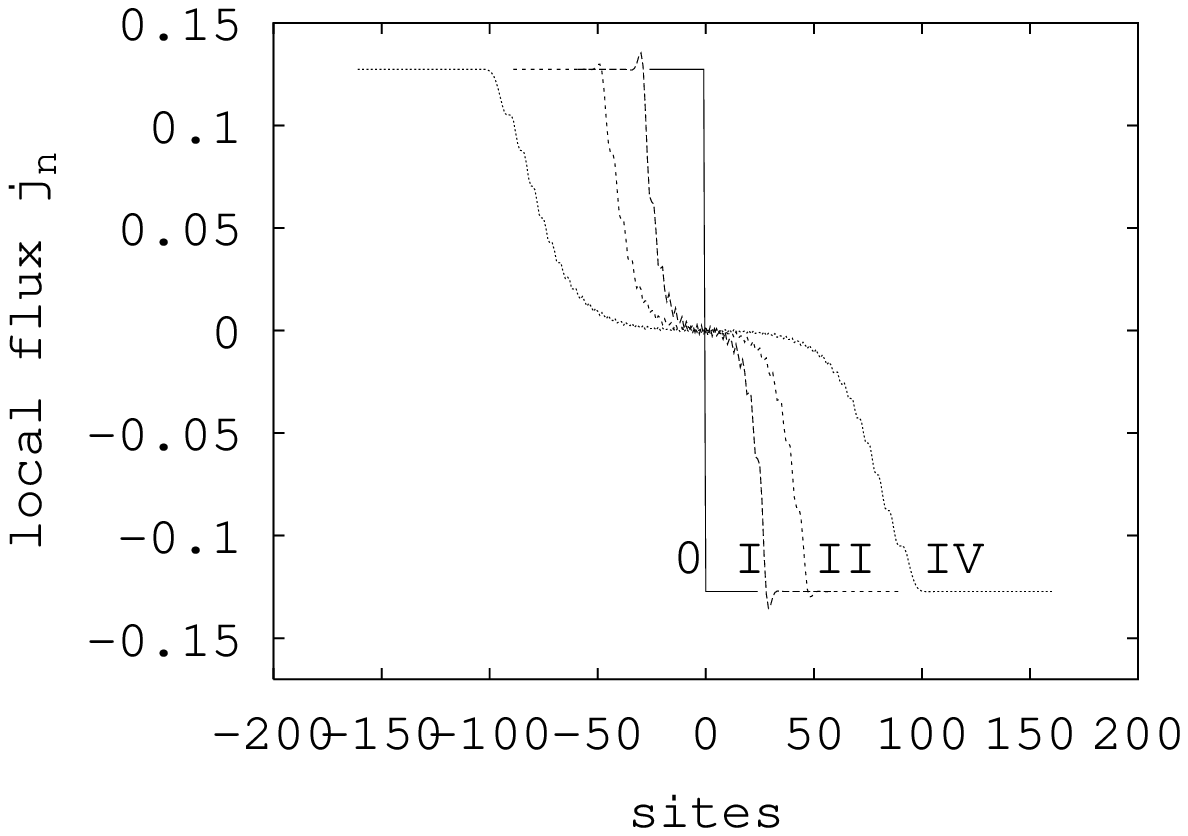}\vspace{3mm}
\caption[]{}
\label{flux_quantum}
\end{figure}

\begin{figure}[htbp]
\setlength{\unitlength}{1.5cm}\vspace{3mm}\hspace*{-2cm}
\epsfig{width=7\unitlength,
%      angle=-90, 
      angle =0,
%      bbllx=250pt, bblly=50pt, bburx=554pt, bbury=770pt,
      file=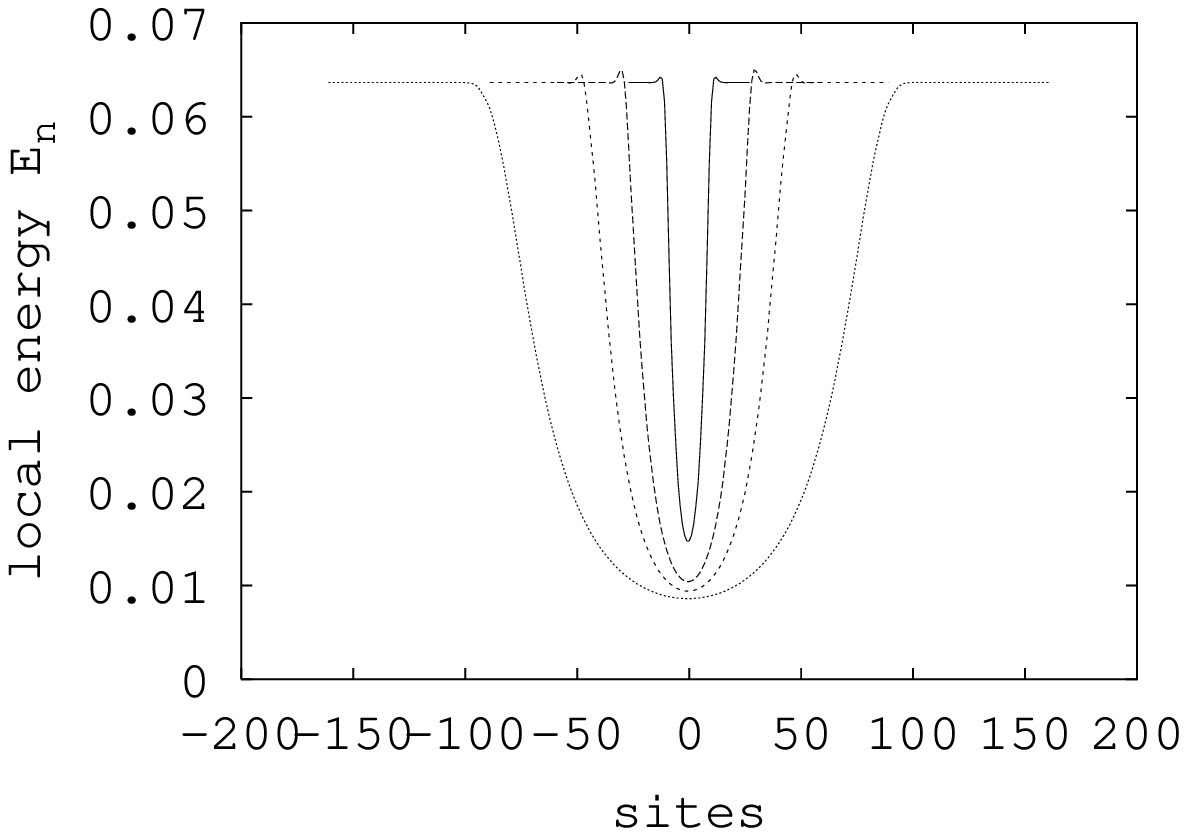}\vspace{3mm}
\caption[]{}
\label{energy_quantum}
\end{figure}


\begin{references}
\bibitem{quantum_computer}  see e.g. Physics World March 1998, Special
issue, pp. 33-57, and references therein.

\bibitem{spinmodel} see e.g. E.H.Lieb and D.Mattis,
"Mathematical Physics in one dimension", Academic Press, NY, (1966),
and references therein.

\bibitem{Mario_hei}  V.V. Konotop, M. Salerno, S. Takeno, {\it Phys.Rev. B}
{\bf 58}, 4892 (1998).

\bibitem{Mario_two_level}  V.V. Konotop, M. Salerno, S. Takeno, {\it %
Phys.Rev. E}{\bf 56}, 7240, (1997).

\bibitem{lieb}  E.H.Lieb, T.Schulz and D. Mattis, {\it Ann. Phys.}{\bf 16}
(1961) 407.

\bibitem{lieb-mattis}  E.H.Lieb and D.Mattis, {\it J. Math. Phys.}{\bf 3},
749 (1962).

\bibitem{jord-wig}  P. Jordan and E. Wigner, {\it Z. Phys.}{\bf 47}
(1928) 631.

\bibitem{Gunter}  T. Antal, Z. R\'{a}cz, A. R\'{a}kos, G. M. Sch\"{u}tz, 
{\it Phys. Rev.E } {\bf 59}, 4912 (1999).

\bibitem{ps}  V. Popkov and M. Salerno, unpublished.

\bibitem{Schotte_privat}  K.D. Schotte, private communication

\bibitem{malomed}  D.J. Kaup, J.El-Reedy, Boris A. Malomed, {\it Phys. Rev.E 
} {\bf 50}, (1994) 1635.
\end{references}
\end{document}